\documentclass[aps,twocolumn,prb]{revtex4-2}

\usepackage{amsmath,graphicx,color}
\usepackage{bm}
\usepackage{physics}

\usepackage{bbold}
\usepackage{lipsum}
\usepackage{natbib}
\usepackage{placeins}
\usepackage{braket}

\newcommand{\FST}{FeSe$_{0.45}$Te$_{0.55}$ }

\newcommand{\FSTT}{FeSe$_{0.45}$Te$_{0.55}$}

\begin{document}
\title{Two-electron photoemission spectroscopy in Topological Superconductors}
\author{Ka Ho Wong$^{1}$, Ameya Patwardhan$^{2,3}$, Peter Abbamonte$^{2,3}$, Fahad Mahmood$^{2,3}$, and Dirk K. Morr$^{1}$}
\affiliation{$^{1}$ Department of Physics, University of Illinois at Chicago, Chicago, IL 60607, USA}
\affiliation{$^{2}$ Department of Physics, University of Illinois at Urbana-Champaign,Urbana,IL 61801, USA}
\affiliation{$^{3}$ Materials Research Laboratory, University of Illinois at Urbana-Champaign,Urbana,IL 61801, USA}

\begin{abstract}
We demonstrate that the photo-electron counting rate, $P^{(2)}$, measured in two electron coincidence spectroscopy (2$e$-ARPES) experiments, provides unprecedented insight into the nature of topological superconductivity. In particular, we show that the spin dependence of $P^{(2)}$ allows one to detect superconducting spin-triplet correlations that are induced in a topological superconductor even in the absence of an associated triplet superconducting order parameter. This ability to detect spin-triplet correlations allows one to distinguish between two recently proposed scenarios for the microscopic origin of topological superconductivity in \FSTT. Finally, we show that $P^{(2)}$ exhibits a characteristic intensity maximum that can be employed to detect topological phase transitions.

\end{abstract}

\maketitle

\section{Introduction}
Topological superconductors harbor Majorana zero modes (MZMs) whose non-Abelian statistics in combination with their topologically protection against disorder and decoherence effects provide an exciting platform for the realization of topological quantum computing \cite{Nayak2008}. However, the experimental observation and identification of MZMs in a variety of superconducting systems \cite{Mourik2012,NadjPerge2014,Das2012,Ruby2015,Pawlak2016,Kim2018,Manna2020,Menard2017,Palacio-Morales2019,Menard2019} met significant challenges due to the system's small superconducting gaps, which are often only of the order of a few hundred $\mu eV$. The recent report of topological superconductivity in the iron-based superconductor \FSTT \cite{Zhang2018,Rameau2019,Zaki2019,Yangmu2021,Wang2018,Machida2019,Kong2019,Zhu2020,Chen2020,Wang2020}, possessing a significantly larger superconducting gap of a few meV, might therefore provide a more suitable platform for the unambiguous identification of MZMs, and the realization of topology based devices and topological quantum computing.

The origin of topological surface superconductivity in \FST was initially proposed to arise from band-inversion  \cite{Wang2015,Wu2016,Xu2016,Zhang2018} -- rendering \FST a 3D topological insulator -- and the gaping of the ensuing surface Dirac cone by proximity induced superconductivity (we refer to this as the 3DTI$^+$ mechanism). However, the recent experimental observation of ferromagnetism on the surface of FeSe$_{1-x}$Te$_{x}$ \cite{Rameau2019,Zaki2019,Yangmu2021,McLaughlin2021} has shed doubts on this interpretation, as topological superconductivity arising from the 3DTI$^+$ mechanism, being protected by a time-reversal symmetry, is destroyed already for rather weak surface ferromagnetism \cite{Wu2021,Xu2021}. A competing scenario was therefore proposed \cite{Mascot2022,Wong2022} in which the very ferromagnetism observed experimentally in combination with the two-dimensional nature of superconductivity in \FST and a Rashba spin-orbit interaction on the surface induced by the broken inversion symmetry, gives rise topological surface superconductivity (we refer to this as the 2DTSC mechanism). Clearly, further experiments are required to distinguish between these two proposed scenarios.

In this article, we demonstrate that the photo-electron counting rate \cite{Berakdar1998}, $P^{(2)}$, measured in two electron coincidence spectroscopy (2$e$-ARPES) experiments, can provide unprecedented insight into the nature of topological superconducting phases, and thus identify the microscopic origin of topological superconductivity in \FSTT. In 2$e$-ARPES experiments, the absorption of a single photon leads to the emission of two coincident photo-electrons. As previously shown \cite{Mahmood2022}, the energy dependence of $P^{(2)}$ cannot only reveal the total center of mass momentum of a Cooper pair, but also its spin state. As a result, 2$e$-ARPES experiments can identify superconducting spin-triplet correlations which are induced within the 2DTSC mechanism in FeSe$_{1-x}$Te$_{x}$, but  are all but absent in the 3DTI$^+$ mechanism.
In addition, we show that 2$e$ARPES experiments can identify topological phase transitions which coincide with a maximum in $P^{(2)}$ for photo-electrons with equal spin. These results open a new venue to distinguish between proposed mechanisms for the emergence of topological superconductivity in FeSe$_{1-x}$Te$_{x}$.

\section{Theoretical Model}
\begin{figure}
\center
\includegraphics[width=8.5cm]{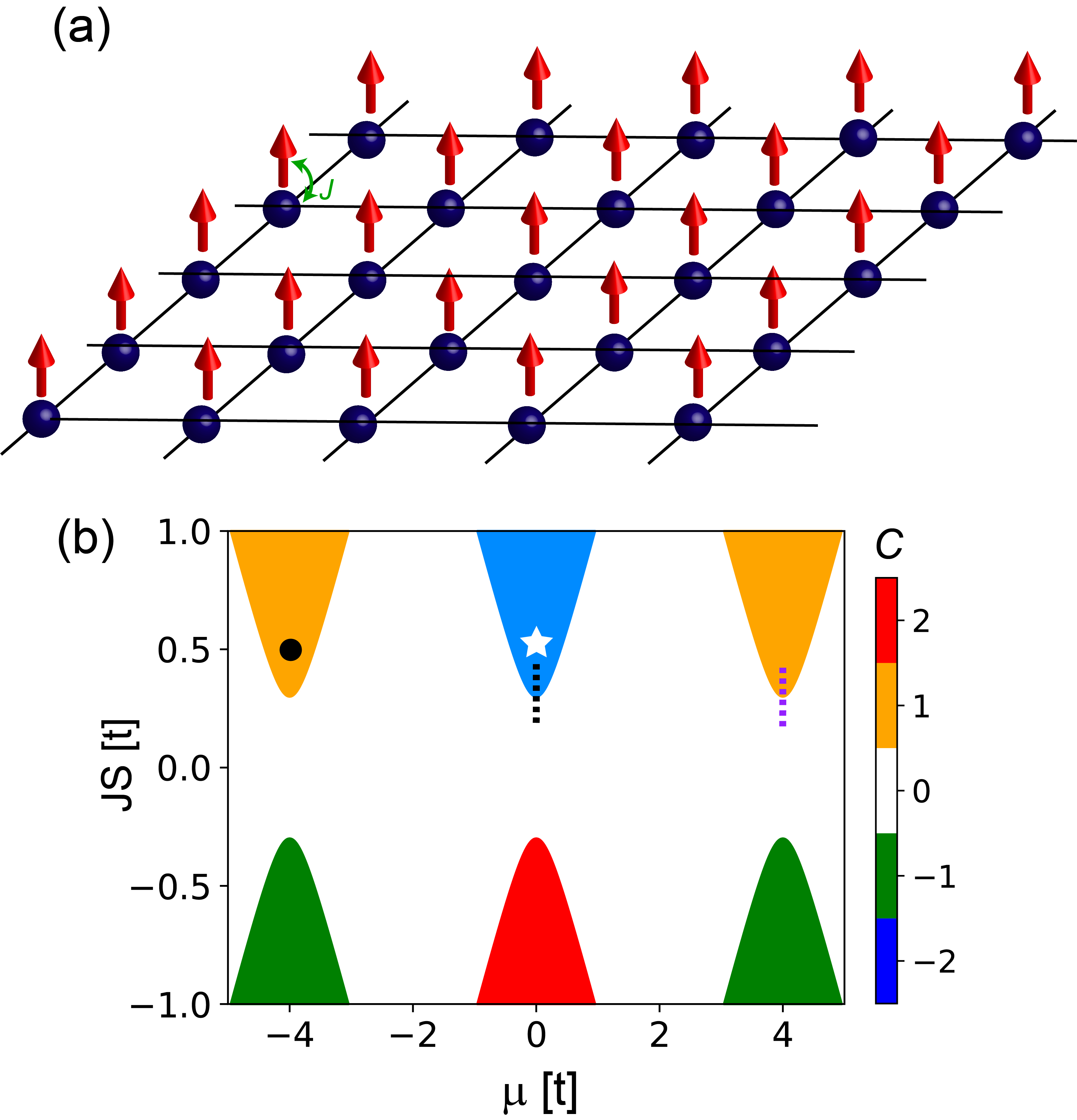}
\caption{(a) Schematic representation of the 2DTSC with local magnetic moments interaction with conduction electrons on the surface of an $s$-wave superconductor. (b) Topological phase diagram of the 2DTSC in terms of the Chern number $C$ in the $(\mu,JS)$-plane. }
\label{fig:Fig1}
\end{figure}

In the following, we consider the 2$e$ARPES photo-electron counting rate for two different types of topological superconductors: (i) a two-dimensional topological superconductor with broken time reversal symmetry, as described by the 2DTSC mechanism, and (ii) a topological superconductor on the surface of a three-dimensional topological insulator, arising from the proximity coupling of its surface Dirac cone to an $s$-wave superconductor, as described by the 3DTI$^+$ mechanism. The former system is described by the Hamiltonian \cite{Rachel2017,Mascot2022}
\begin{align}
    H_{SC}&=\sum_{\vb{k}}\Bigg[\xi_{\vb{k}}c^{\dagger}_{\vb{k},\sigma}c_{\vb{k},\sigma}+\Delta_{0}\qty(c^{\dagger}_{\vb{k},\uparrow}c^{\dagger}_{-\vb{k},\downarrow}+c_{\vb{k},\downarrow}c_{-\vb{k},\uparrow})\nonumber\\
    &\qquad\qquad +2\alpha\sum_{\pmb{\delta},\sigma,\sigma'}\sin(\vb{k}\cdot\pmb{\delta})c^{\dagger}_{\vb{k},\sigma}\qty(\pmb{\delta}\times\pmb{\sigma})^{z}_{\sigma\sigma'}c_{\vb{k},\sigma'}\nonumber\\
     &\qquad\qquad-JS\sum_{\sigma,\sigma'}c^{\dagger}_{\vb{k},\sigma}\sigma^{z}_{\sigma\sigma'}c_{\vb{k},\sigma'}\Bigg] \, .
\end{align}
Here, $c^{\dagger}_{\vb{k},\sigma}$ creates an electron with momentum ${\bf k}$ and spin $\sigma$ and $\xi_{\vb{k}}=-2t\qty(\cos{k_x}+\cos{k_y})-\mu$ is the tight binding dispersion with $-t$ being the nearest-neighbor hopping amplitude, and  $\mu$ being the chemical potential. Moreover, $\Delta_0$ is the s-wave superconducting order parameter, $\alpha$ is the Rashba spin-orbit interaction, with $\pmb{\delta}$ being the unit vector connecting nearest neighbor sites, and $J$ is the magnetic exchange coupling between the ordered moments of magnitude $S$ and the conduction electrons. A schematic representation of the 2DTSC is shown in Fig.~\ref{fig:Fig1}(a). This topological superconductor is in the topological class D, and its  topological invariant is the Chern number \cite{Rachel2017}, $C$. Its topological phase diagram in terms of $C$ is shown in Fig.~\ref{fig:Fig1}(b) in the $(\mu,JS)$-plane.

The Hamiltonian of the 3DTI system is given by  \cite{Schubert2012}
\begin{eqnarray}
H_{3D}= &-& t\sum_{{\bf r},j=1,2,3} \left(\Psi_{{\bf r}+\hat{e}_j}^{\dagger}  \frac{\Gamma^1-i\Gamma^{j+1}}{2} \Psi_{\bf r} \nonumber+H.c.\right) \\
&+& m \sum_{\bf r} \Psi_{\bf r}^\dagger \Gamma^1 \Psi_{\bf r}
\end{eqnarray}%
with spinor
\begin{align}
    \Psi_{{\bf r}}^{\dagger} = \left( c_{{\bf r},1,\uparrow}^{\dagger}, c_{{\bf r},2,\uparrow}^{\dagger}, c_{{\bf r},1,\downarrow}^{\dagger}, c_{{\bf r},2,\downarrow}^{\dagger} \right) \, ,
\end{align}
where $c_{{\bf r},a,\sigma}$ annihilates an electron with spin $\sigma$ in orbital $a=1,2$ at site ${\bf r}$, and $\Gamma^{(0,1,2,3,4)} = (\mathbb{1} \otimes \mathbb{1}, \mathbb{1} \otimes s_z, -\sigma_y \otimes s_x, \sigma_x \otimes s_x, -\mathbb{1} \otimes s_y)$ with $\sigma_i$ and $s_i (i=x,y,z)$ being Pauli matrices. Within this model, a topological superconducting phase emerges on the surface of the 3DTI due to proximity coupling to a superconductor, and the ensuing opening of a gap in the 3DTI's surface Dirac cone. The proximity induced superconductivity is described by the Hamiltonian
\begin{align}
H_{\Delta} = \Delta_0 \sum_{\mathbf{r},a=1,2}c^{\dagger}_{\mathbf{r},a,\uparrow}c^{\dagger}_{\mathbf{r},a,\downarrow}+H.c. \ ,
\end{align}
where $\Delta_0$ is the induced superconducting order parameter with $s$-wave symmetry.
We note that to demonstrate the qualitative different form of $P^{(2)}$ in the 2DTSC and 3DTI$^+$ systems (see below), it is sufficient to consider for simplicity an $s$-wave symmetry of the superconducting order parameter, rather than the $s_{\pm}$-symmetry found in \FSTT. A more material specific calculation of $P^{(2)}$ that also takes into account the multi-band structure of \FST \cite{Mascot2022}, is reserved for a future study.

In 2$e$ARPES spectroscopy, there are two distinct processes in which the absorption of a single photon leads to the ejection of a correlated pair of electrons  \cite{Haak1978,Berakdar1998,Chiang2020,Fominykh2002,Schumann2007,Schumann2011,Schumann2012,Herrmann1998,Mahmood2022}, giving rise to the photo-electron counting rate $P^{(2)}$. In the first one, the absorption of a photon results in the excitation of a valence band electron into a free photo-electron state, which subsequently ejects a second valence electron via an electron energy-loss (EELS)-like scattering event, mediated by the Coulomb interaction. In the second process, the photon first excites a photo-electron from a core-level (rather than the conduction band). The resulting core hole is then filled by a valence electron, leading to the emission of a second valence electron through an Auger process. As previously shown \cite{Mahmood2022}, both processes lead to a very similar energy, momentum and spin dependence of $P^{(2)}$. However, the use of lower photon energy, laser based XUV sources will not allow 2$e$-ARPES experiments to directly probe core states. Below, we will therefore consider $P^{(2)}$ as arising from the first process only, which is described by the Hamiltonian
\begin{align}
    H_{scat} & = \sum_{{\bf k,q},\sigma,\nu} \gamma_{\nu}({\bf q}) d^\dagger_{{\bf k+q},\sigma} c_{{\bf k},\sigma} \left( a_{{\bf q},\nu}  + a^\dagger_{-{\bf q},\nu}\right)  \nonumber \\
    & + \sum_{{\bf k,p,q},\alpha, \beta} V({\bf q}) d^\dagger_{{\bf k}+{\bf q},\alpha} d^\dagger_{{\bf p}-{\bf q},\beta} d_{{\bf p},\beta} c_{{\bf k},\alpha}  + h.c.
    \label{eq:Ham}
\end{align}
Here, $\gamma_{\nu}({\bf q})$ is the effective electron-photon dipole interaction, $d^\dagger_{{\bf k},\sigma} (c_{{\bf k},\sigma})$ creates (destroys) a photo-electron (conduction electron) with momentum ${\bf k}$ and spin $\sigma$, and $V({\bf q}) = V_0/ \left({\bf q}^2 + \kappa^{2}\right)$ is the Fourier transform of the (screened) Coulomb interaction, with $\kappa^{-1}$ being the screening length. We note that in the 3DTI, possessing two orbitals per site, the $c$-electron operators in Eq.(\ref{eq:Ham}) acquire an orbital index. Since the qualitative nature of $P^{(2)}$ does not depend on $\kappa$ \cite{Mahmood2022}, we take for concreteness $\kappa^{-1} = 10 a_0$ for the results shown below. Moreover, as the photon momentum is much smaller than typical fermionic momenta, we set it equal to zero, such that $\gamma_{\nu}({\bf q}) = \gamma_0$ is simply a momentum-independent constant.

To compute the photo-electron counting rate in 2$e$ARPES experiments, which depends on the two photo-electron momenta and spin projections, we use
\begin{equation}
 P^{(2)}({\bf k}_1^\prime, \sigma^\prime_1, {\bf k}_2^\prime, \sigma^\prime_2 ) = \frac{1}{Z} \sum_{a,b} \frac{e^{-\beta E_a}}{\Delta t} \left| \langle  \Psi_b | {\hat S}^{(2)}(\infty,-\infty) | \Psi_a \rangle \right|^2
 \label{eq:P2a}
\end{equation}
where $Z$ is the partition function, and the initial and final states of the entire system before and after the two-step process, $\ket{\Psi_a}$ and $\ket{\Psi_b}$, are represented by the following product states
\begin{align}
    \ket{\Psi_a}&=\ket{\Phi_{a}}\ket{1_{\vb{q},\nu}}_{p}\ket{0}_{pe}\nonumber\\
    \ket{\Psi_b}&=\ket{\Phi_{b}}\ket{0}_{p}\ket{1_{\vb{k}_1',\sigma_1'} 1_{\vb{k}_2',\sigma_2'}}_{pe}
\end{align}
where $\ket{1_{\vb{q},\nu}}_{p}$ is the initial state of the photon with wavevector $\vb{q}$ and polarization $\nu$, and $\ket{1_{\vb{k}_1',\sigma_1'} 1_{\vb{k}_2',\sigma_2'}}_{pe}$ is the final photo-electron state with momenta $\vb{k}_1',\vb{k}_2'$ and spins $\sigma_1,\sigma_2$.
\begin{figure}[htb]
\center
\includegraphics[width=\linewidth]{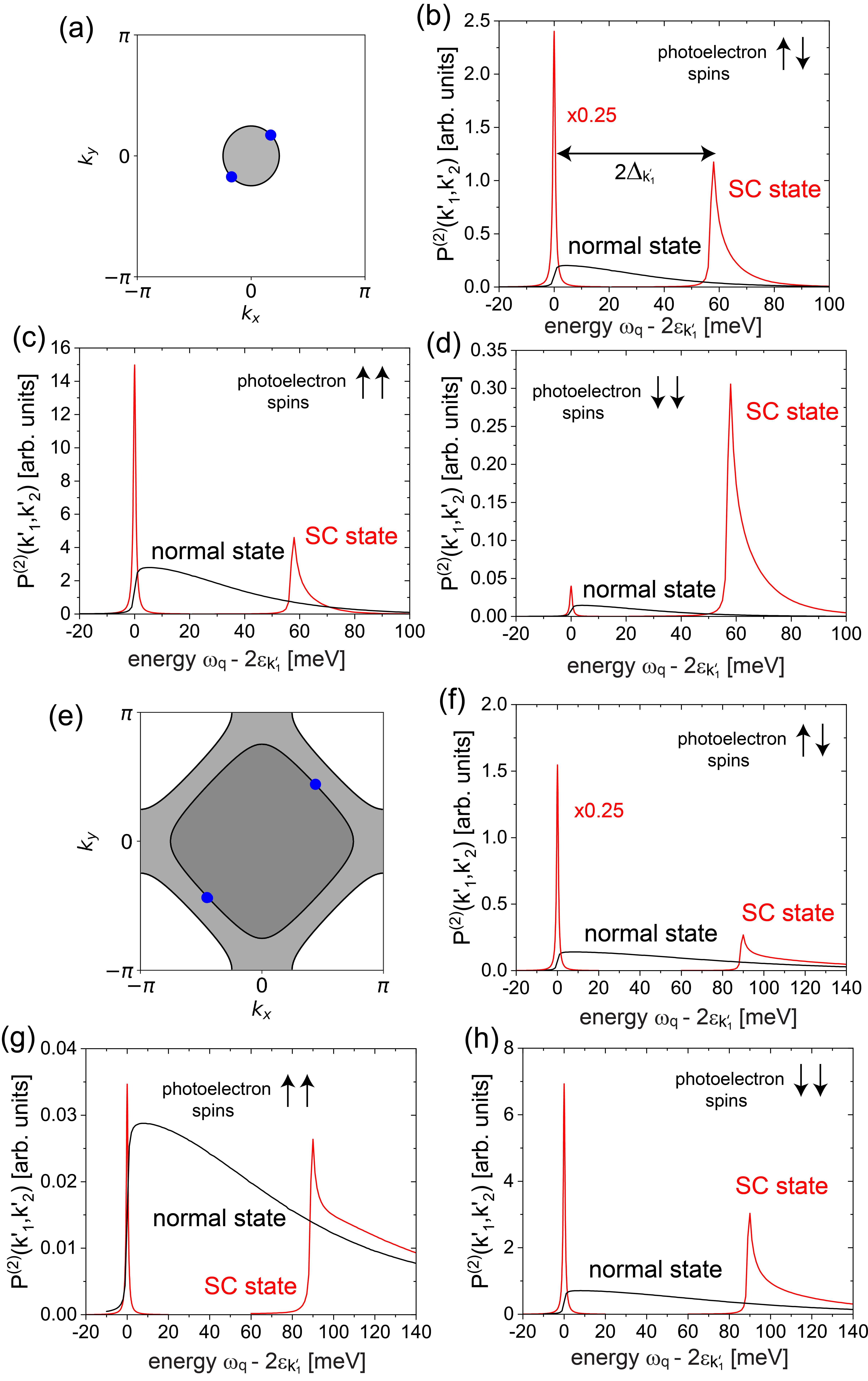}
\caption{$P^{(2)}$ in the topological $C=1$ phase at $\mu=-4t$ [black dot in Fig.~\ref{fig:Fig1}(b)]: (a) Fermi surface in the normal state, and  (b)-(d) the corresponding $P^{(2)}$ for three different spin configurations of the photo-electrons, in the normal (black lines) and superconducting (red lines) states. The photo-electron momenta ${\bf k}_{1,2}^\prime$ are shown as filled blue circles in (a). $P^{(2)}$ in the topological $C=-2$ phase at $\mu=0$ [white star in Fig.~\ref{fig:Fig1}(b)]: (e) Fermi surface in the normal state, and  (f)-(h) the corresponding $P^{(2)}$ for three different spin configurations of the photo-electrons , in the normal (black lines) and superconducting (red lines) states. The photo-electron momenta ${\bf k}_{1,2}^\prime$ are shown as filled blue circles in (e). Parameters are $t=200$meV and $(\alpha,\Delta_0,JS)=(0.2,0.3,0.5)$t \cite{Rachel2017}. }
\label{fig:Fig2}
\end{figure}
The sum in Eq.(\ref{eq:P2a}) runs over all states $|\Phi_{a,b} \rangle$ of the topological superconductor, $\Delta t$ is the time over which the photon beam is incident in the superconductor, and ${\hat S}^{(2)}$ is the second-order contribution to the $S$-matrix arising from $H_{scat}$.
As previously discussed \cite{Mahmood2022}, the photo-electron counting rate can then be written as $P^{(2)}=V^2 P^{(2)}_{SC}+V P^{(2)}_{2cp}$, where $V$ is the volume of the system. The first term arises from the breaking of a single Cooper pair, and thus directly reflects the strength of the superconducting condensate, while the second term arises from the breaking of two Cooper pairs. Note that the first term scales as $V^2$, while the second term scales as $V$, as the probability to find a second Cooper pair from which an electron is ejected is given by $1/V$.

\section{Results}

\subsection{$P^{(2)}$ in a 2DTSC}
We begin by considering the photo-electron counting rate in the topological $C=1$ phase of the 2DTSC system [at $\mu=-4t$, see black circle in Fig.~\ref{fig:Fig1}(b)], whose the Fermi surface in the normal state is shown in Fig.~\ref{fig:Fig2}(a). $P^{(2)}$ for two photo-electrons with opposite momenta (${\bf k}_1^\prime =- {\bf k}_2^\prime$) located on the Fermi surface [see filled blue circles in Fig.~\ref{fig:Fig2}(a)], and opposite spins, is shown in Fig.~\ref{fig:Fig2}(b), both for the normal and superconducting states. In the normal state, the onset of $P^{(2)}$ occurs when the energy of the photon is sufficiently large to eject two photo-electrons with energy $\varepsilon_{{\bf k}_1^\prime}$ (which also includes the work function). In the superconducting state, $P^{(2)}$ exhibits two contributions. The first one is a peak at $\Delta \omega = \omega_{\bf q}-2\varepsilon_{{\bf k}^\prime_1}=0$ which arises from $P^{(2)}_{SC}$ and reflects the breaking of a single Cooper pair. This peak is present only if the two photo-electrons possess the same center-of-mass momentum and spin state as a Copper pair \cite{Mahmood2022}, as is the case here. The second contribution, arising from $P^{(2)}_{2cp}$, is a continuum with onset energy $\Delta \omega \approx 2 \Delta_{{\bf k}^\prime_1}$, with $\Delta_{{\bf k}_1'}$ being the superconducting gap at ${\bf k}_1'$, implying that the two measured photo-electrons are ejected from two different Cooper pairs. As an essential feature of a 2DTSC is the emergence of spin-triplet correlations \cite{Rachel2017}, we present $P^{(2)}$ for two photo-electrons possessing equal spins in Figs.~\ref{fig:Fig2}(c),(d). Interestingly enough, we find that even for this spin configuration, a peak at $\Delta \omega = 0$ exists, implying the presence of spin-triplet Cooper pairs.  As the starting Hamiltonian, Eq.(\ref{eq:Ham}), does not contain a triplet superconducting order parameter, this peak at $\Delta \omega = 0$ should be attributed to the presence of superconducting spin-triplet correlations, which are induced through the combination of ferromagnetism, Rashba spin-orbit interaction, and $s$-wave superconductivity.

To gain more insight into the relation between $P^{(2)}_{SC}$ and the superconducting pairing correlations, we consider the latter in the spin triplet channel ($S=1, S_z=\pm 1$), given by  $C_{T,\sigma}=\langle c^\dagger_{{\bf k},\sigma} c^\dagger_{-{\bf k},\sigma} \rangle$  and $\sigma = \uparrow, \downarrow$, and singlet channel ($S=0$), given by  $C_S=\langle c^\dagger_{{\bf k},\uparrow} c^\dagger_{-{\bf k},\downarrow} - c^\dagger_{{\bf k},\downarrow} c^\dagger_{-{\bf k},\uparrow} \rangle$. We note that the superconducting correlations in the spin triplet channel $S=1, S_z=0$, are identically zero, such that $P^{(2)}$ for photo-electrons with opposite spins arises solely from the presence of superconducting correlations in the singlet channel. For the same momentum on the Fermi surface [see filled blue circles in Fig.~\ref{fig:Fig2}(a)] as we considered for the calculation of $P^{(2)}$ in Figs.~\ref{fig:Fig2}(b)-(d), we obtain $C_{T,\uparrow} \approx 0.347$, $C_{T,\downarrow} \approx 0.026$, and $C_S=0.252$. Given that $P^{(2)}_{SC} \sim C^2$ \cite{Mahmood2022}, the relative peak intensities in $P^{(2)}_{SC}$ in the three different channels are in good agreement with the relative strength of the superconducting correlations. This supports our conclusion that $P^{(2)}_{SC}$ indeed reflects the superconducting correlations in the system.  Finally, we note that we obtain qualitatively similar results for $P^{(2)}$ in the topological $C=-2$ phase of the 2DTSC [at $\mu=0$, see white star in Fig.~\ref{fig:Fig1}(b)], whose Fermi surface in the normal state is shown in Fig.~\ref{fig:Fig2}(e).  $P^{(2)}$ shown in Figs.~\ref{fig:Fig2}(f)-(h)  exhibits a peak at $\Delta \omega = 0$ for all three spin configurations, again reflecting the presence of spin-triplet Cooper pairs.

While the presence of spin-triplet correlations is a key feature of a 2DTSC, as shown in Fig.~\ref{fig:Fig2}, it is not sufficient to uniquely identify a topological phase, as the interplay of ferromagnetism, a Rashba spin-orbit coupling, and an $s$-wave superconducting order parameter can also induce spin-triplet correlations in a trivial superconducting phase \cite{Rachel2017}. The question thus arises whether 2$e$-ARPES experiments can be employed to distinguish between topological and trivial superconducting phases, or to at least identify topological phase transitions.
\begin{figure}[htb]
\center
\includegraphics[width=\linewidth]{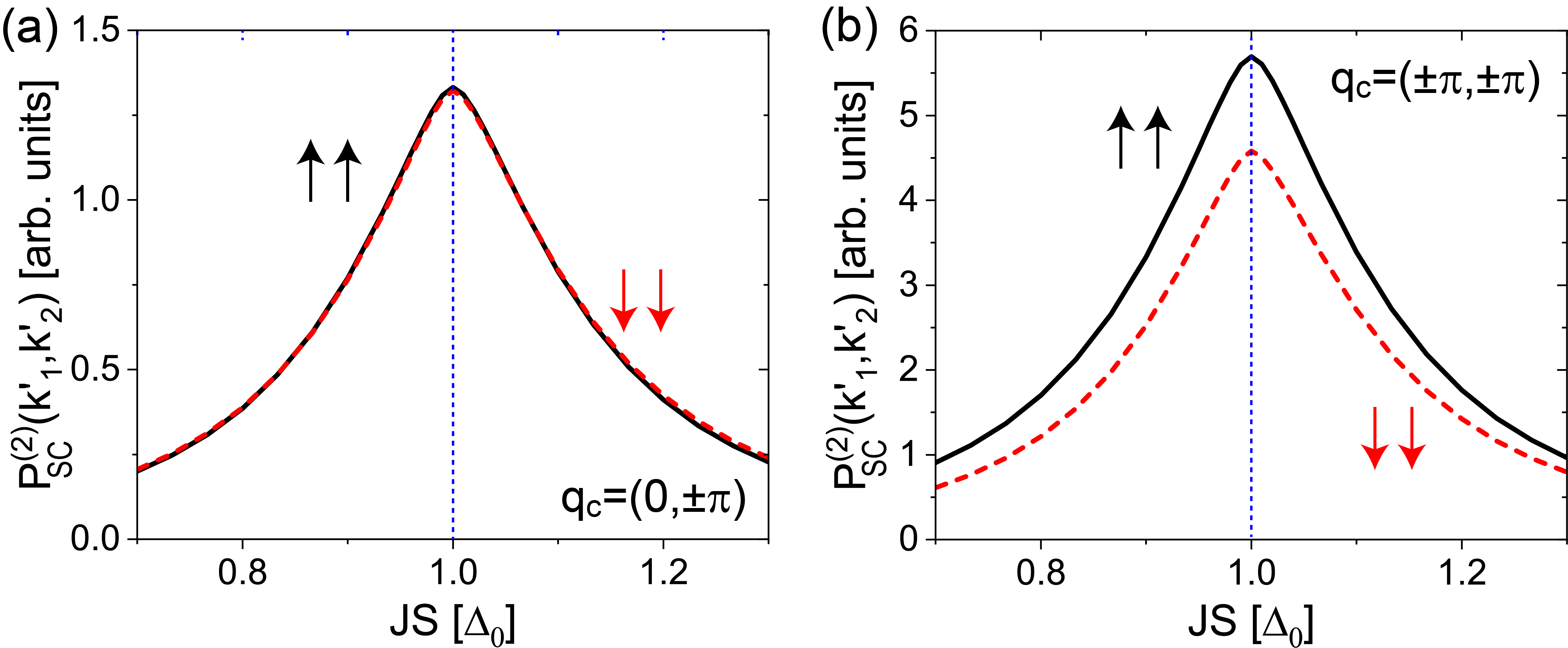}
\caption{$P^{2}_{SC}$ at $\Delta\omega=0$ for photo-electrons with equal spin projections and momenta ${\bf k}_1^\prime = -{\bf k}_2^\prime ={\bf q_c}$ at the gap closing momentum ${\bf q_c}$, across a topological phase transition into the (a) $C=-2$ [$\mu=0, {\bf q}_c = (0,\pm \pi), (\pm \pi,0)$], and (b) $C=1$ [$\mu=4t, {\bf q}_c = (\pm \pi,\pm \pi)$] phases. The vertical dashed blue line indicates the topological phase transition at $(JS)_{cr}=\Delta_0$. Parameters are $t=200$meV and $(\alpha,\Delta_0)=(0.2,0.3)t$. }
\label{fig:Fig3}
\end{figure}
To answer this question, we consider the topological phase transition between a trivial phase and the topological $C=-2$ phase, along the black dashed line shown in Fig.~\ref{fig:Fig1}(b), which is accompanied by the closing of the superconducting gap at the $X/Y$-points, yielding a gap closing momentum, ${\bf q}_c = (0,\pm \pi), (\pm \pi,0)$. We thus consider two photo-electrons with momenta ${\bf k}_1^\prime = - {\bf k}_2^\prime={\bf q}_c$ and present in Fig.~\ref{fig:Fig3}(a), $P^{(2)}_{SC}$ at $\Delta \omega =0$ as a function of $JS$. As we tune the system through a topological phase transition at $(JS)_{cr}=\Delta_0$, we find that while $P^{(2)}_{SC}$ is non-zero both in the topological and trivial phases, consistent with earlier findings \cite{Rachel2017}, it exhibits a maximum at $(JS)_{cr}$. We obtain the same result for the phase transition between a trivial phase and the topological $C=1$ phase as shown in Fig.~\ref{fig:Fig3}(b), along the purple dashed line in Fig.~\ref{fig:Fig1}(b), where the gap closing occurs at ${\bf q}_c = (\pm \pi,\pm \pi)$. The fact that $P^{(2)}_{SC}$ exhibits a maximum at $(JS)_{cr}$ for both cases implies that this maximum is a general signature of topological phase transitions in a 2DTSC, that can be employed to identify them.

\begin{figure}[htb]
\center
\includegraphics[width=8.cm]{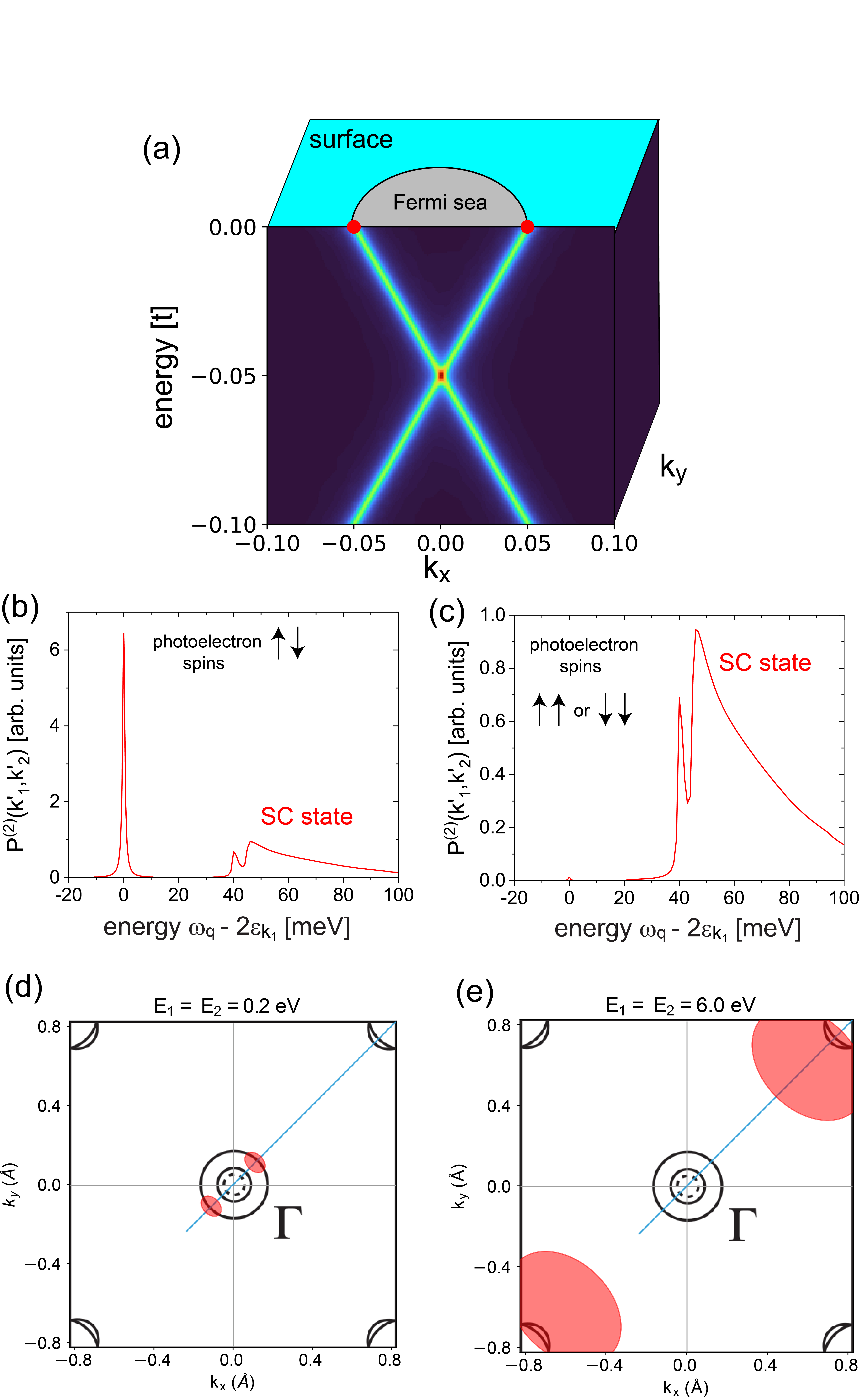}
\caption{(a) Schematic representation of a surface Dirac cone in a 3DTI with $N_z=5$ layers. (b),(c) $P^{(2)}$ for photo-electrons with momenta indicated by filled red circles in (a) and different spin states. Parameters used here are $t=200$meV and $(m,\mu,\Delta_0)=(2,0.2,0.3)t$.(d)-(e) Schematic form of the Fermi surface of FeSe$_{0.5}$Te$_{0.5}$ with calculated projections of fields of view (red circles) of two time-of-flight (TOF) analysers separated by $90^{\circ}$. (d) Calculation done with 0.2 eV kinetic energy of the two electrons (or an incident photon energy of $h\nu \approx 9\textnormal{eV}$), corresponding to electrons from the hole-like Fermi Surface. (e) Calculation done with 6.0 eV kinetic energy of the two electrons (or an incident photon energy of $h\nu \approx 21 \textnormal{eV}$), corresponding to electrons from the electron-like Fermi surface. (d) and (e) adapted from \cite{Zhang2018}.
}
\label{fig:Fig5}
\end{figure}

\subsection{$P^{(2)}$ in a 3DTI$^{+}$ system}

Finally, we consider $P^{(2)}$ in the topological superconductor arising from the 3DTI$^+$ mechanism whose surface Dirac cone in the normal state is schematically shown in  Fig.~\ref{fig:Fig5}(a). To this end, we consider a system that is translationally invariant in the $x$- and $y$-directions and possesses a finite number of layers $N_z$ in the $z$-direction. We assume that photo-electrons are ejected from the surface layer (which contains the Dirac cone) only, as the spectral weight of the Dirac cone on layers below the surface is negligible. Moreover, we assume below that both photo-electrons of the same orbital character in the 3DTI$^+$ system (orbital selectivity can in general be achieved by varying the energy or polarization of the incoming photon \cite{Sobota2021}). However, $P^{(2)}$ is independent of the orbital from which the photo-electrons are ejected.

In Fig.~\ref{fig:Fig5}(b), we present $P^{(2)}$ for photo-electrons ejected from the same orbital with opposite spins and momenta ${\bf k}_1^\prime = - {\bf k}_2^\prime$ on the Fermi surface of the Dirac cone [see filled blue circles in Fig.~\ref{fig:Fig5}(a)].
As expected, $P^{(2)}$  exhibits a peak at $\Delta\omega=0$ arising from the breaking of a single (singlet) Cooper pair, and an onset of a continuum at $\Delta\omega \approx 2 \Delta_{{\bf k}_1^\prime}$ arising from breaking of 2 Cooper pairs. In contrast, for photo-electrons with the same spin state, $P^{(2)}$ exhibits a peak at $\Delta\omega=0$ whose intensity is more than 500 times smaller [see Fig.~\ref{fig:Fig5}(c)] than that for the case of opposite photo-electron spins [see Fig.~\ref{fig:Fig5}(b)]. This is consistent with vanishingly small superconducting correlations in the triplet $S=0, S_z=\pm 1$ channel (as before, the correlations in the $S=0, S_z=0$ channel are identically zero). The reason that the triplet correlations are vanishingly small is  that the surface Dirac cone possesses a helical structure, reflecting spin-momentum locking, which implies that states with opposite momenta ${\bf k}_1^\prime = -{\bf k}_2^\prime$ possess an opposite spin polarization. As such, these states can only form Cooper pairs in the singlet channel. However, this spin-momentum locking, and the associated complete suppression of pairing in the spin-triplet channel, is lifted when the Dirac point is located away from zero-energy at non-zero $E_D$. However, since the degree to which the spin-momentum locking is violated scales with $E_D$, the spin-triplet correlations and the associated intensities in $P^{(2)}_{SC}$ remain very small for realistic values of $E_D$  \cite{Zhang2018,Rameau2019,Zaki2019,Yangmu2021}, as shown in Fig.~\ref{fig:Fig5}(c).
Thus, in the 3DTI$^+$ system, the $\Delta \omega = 0$ peak in $P_{SC}^{(2)}$ for equal spin configuration is either absent, or at least greatly suppressed in comparison to the peak in $P_{SC}^{(2)}$ for opposite spin projections, in contrast to the results in the 2DTSC system (see Fig.~\ref{fig:Fig2}). We thus conclude that the presence or absence of a $\Delta\omega=0$ peak in $P^{(2)}$ for photo-electrons with equal spin states is a characteristic signature of topological superconductivity arising from the 2DTSC and 3DTI$^+$ mechanisms, respectively, and thus allows us to discriminate between them.
We note that our findings are directly applicable to the case of \FST to distinguish between the recently proposed 2DTSC \cite{Mascot2022,Wong2022} and 3DTI$^+$ \cite{Wang2015,Wu2016,Xu2016,Zhang2018} mechanisms for the emergence of topological surface superconductivity. In this case, 2eARPES can explicitly probe both the hole- and electron-like Fermi surfaces using two time-of-flight (TOF) analyzers and XUV photon energies of $9\textnormal{eV}$ and $21\textnormal{eV}$ (as illustrated in Fig.~\ref{fig:Fig5}(d) and Fig.~\ref{fig:Fig5}(e). These experimental requirements are well within the scope of current 2eARPES instruments \cite{Chiang2020}

\section{Conclusions}
We have demonstrated that 2$e$ARPES experiments can detect superconducting spin-triplet correlations that are induced in a 2DTSC, even in the absence of an associated long-range order parameter. This allows us to identify characteristic signatures in the 2$e$ARPES photo-electron counting rate, $P^{(2)}$ for equal spin configuration that distinguish between the recently proposed 2DTSC \cite{Mascot2022,Wong2022} and 3DTI$^+$ \cite{Wang2015,Wu2016,Xu2016,Zhang2018} mechanisms for the emergence of topological surface superconductivity in \FSTT. Finally, we showed that $P_{SC}^{(2)}$ exhibits a characteristic feature -- a maximum in intensity -- at a topological phase transition that allows one to identify its occurrence. These results show that 2$e$ARPES spectroscopy represents an invaluable experimental probe in the study of topological superconductors.

\section{Acknowledgements}
This study was supported by the Center for Quantum Sensing and Quantum Materials, an Energy Frontier Research Center funded by the U. S. Department of Energy, Office of Science, Basic Energy Sciences under Award DE-SC0021238.
F.M. acknowledges support from the EPiQS program of the Gordon and Betty Moore Foundation, Grant GBMF11069.

%

\end{document}